\documentclass[prd,floatfix,nofootinbib,twocolumn]{revtex4-1}

\usepackage{exscale}                  % correct scaling of math-symbols
\usepackage[intlimits]{amsmath}       % improved mathematical typesetting
\usepackage{amsfonts,bbm}
\usepackage{amssymb,amscd}
\usepackage[dvips]{epsfig}
\usepackage{graphicx}
\usepackage{array}
\usepackage{bm}
\usepackage{nicefrac}
\usepackage[colorlinks=true,linkcolor=blue,citecolor=blue,urlcolor=blue]{hyperref}

\usepackage{subfigure}  % use for side-by-side figures

\newcolumntype{L}[1]{>{\raggedright\let\newline\\\arraybackslash\hspace{0pt}}m{#1}}
\newcolumntype{C}[1]{>{\centering\let\newline\\\arraybackslash\hspace{0pt}}m{#1}}
\newcolumntype{R}[1]{>{\raggedleft\let\newline\\\arraybackslash\hspace{0pt}}m{#1}}
\usepackage{color}
\newcommand{\is}{\sum\!\!\!\!\!\!\!\int}

\newcommand{\beqa}{\begin{eqnarray}}
\newcommand{\eeqa}{\end{eqnarray}}
\newcommand{\beq}{\begin{equation}}
\newcommand{\eeq}{\end{equation}}

\newcommand{\Tr}{\mathrm{Tr}}
\newcommand{\vect}[1]{\bm{#1}}
\newcommand{\conjg}[1]{\ensuremath{\hspace{1pt}\overline{\hspace{-1pt}#1\hspace{-1pt}}}\hspace{1pt}}

\definecolor{myred}{rgb}{1,0.8,0.8}

 \newcommand{\twopartdef}[4]
{
	\left\{
		\begin{array}{ll}
			#1 & \mbox{if } #2 \\
			#3 & \mbox{if } #4
		\end{array}
	\right.
}

\begin{document}
\title{Baryon effects on the location of QCD's critical end point}

\author{Gernot Eichmann}
\author{Christian~S.~Fischer}
\author{Christian~A.~Welzbacher}

\affiliation{Institut f\"ur Theoretische Physik,
Justus-Liebig-Universit\"at Gie\ss{}en, Heinrich-Buff-Ring 16, D-35392 Gie\ss{}en, Germany.}

\date{\today}

\begin{abstract}
The location of the critical end point of QCD has been determined in previous
studies of $N_f=2+1$ and $N_f=2+1+1$ dynamical quark flavors using a (truncated)
set of Dyson-Schwinger equations for the quark and gluon propagators of Landau-gauge
QCD. A source for systematic errors in these calculations has been the
omission of terms in the quark-gluon interaction that can be parametrized in
terms of baryonic degrees of freedom. These have a potentially large dependence
on chemical potential and therefore may affect the location of the critical end
point. In this exploratory study we estimate the effects of these contributions,
both in the vacuum and at finite temperature and chemical potential. We find only
a small influence of baryonic contributions on the location of the critical end
point. We estimate the robustness of this result by parameterizing further
dependencies on chemical potential.
\end{abstract}

\maketitle

\section{Introduction}
Heavy ion collision experiments at RHIC/BNL and the future CBM/FAIR facility probe
the phase structure of QCD at finite chemical potential. One of the major goals of
these experiments is the study of the existence, the location and the properties of
a critical end point (CEP), where the chiral crossover at small chemical potential turns
into a first-order transition.

From a theoretical point of view it remains unclear at the moment whether this is
indeed the case. Lattice calculations firmly established the crossover behavior
at zero chemical potential, see e.g. \cite{Borsanyi:2010bp,Bazavov:2011nk} and references
therein. At finite chemical potential, lattice calculations are hampered by the
notorious fermion sign problem. Although various extrapolation methods agree with
each other at rather small chemical potential
\cite{Fodor:2001pe,Gavai:2004sd,deForcrand:2010he,Kaczmarek:2011zz,Endrodi:2011gv,
Cea:2012ev,Bonati:2015bha,Bellwied:2015rza,Cea:2015cya},
for regions in the $(T,\mu)$ plane with $\mu_B/T > 2$ uncertainties accumulate rapidly.
Therefore, despite many efforts the basic properties of the phase diagram of QCD are not
yet settled, see e.g. \cite{Weise:2012yv,Fukushima:2011jc} and references therein. Thus other
theoretical methods are mandatory to complement the lattice calculations.

In a sequence of previous papers \cite{Fischer:2012vc,Fischer:2013eca,Fischer:2014ata}
the location of the CEP for $N_f=2$, $N_f=2+1$ and $N_f=2+1+1$ quark
flavors has been determined using the (truncated) Dyson-Schwinger equations (DSEs) for the
quark and gluon propagators of QCD. A particular focus in these studies has been the
inclusion of back-reaction effects of the quarks onto the Yang-Mills sector, which
allowed to go beyond simple modeling of the gluon part either within the DSE approach
\cite{Qin:2010nq,Wang:2014yla} or in chiral models like the Nambu-Jona-Lasinio (NJL) 
model~\cite{Berges:1998rc}, its Polyakov-loop extended versions~\cite{Fukushima:2003fw,Megias:2004hj,Ratti:2005jh}
and the Polyakov-loop extended quark-meson (PQM) model~\cite{Schaefer:2007pw,Skokov:2010wb,Herbst:2010rf}.
At zero chemical potential the inclusion of the quark back-reaction on the gluons produced
the correct temperature behavior of the quark condensate and led to predictions for
the magnitude of unquenching effects in the gluon propagator \cite{Fischer:2012vc}, which
have been verified by subsequent lattice calculations \cite{Aouane:2012bk}.

In this approach a CEP has been found at rather large quark chemical
potential $(T^c,\mu_q^c)=(115,168) \,\mbox{MeV}$. Since this result relies on a
truncation of the quark-gluon interaction which is still far from complete, it is an
important task to quantify its systematic error. In particular, effects due to a nonzero
chemical potential that cannot be tested by comparison with lattice calculations
at $\mu_q=0$ may provide for sizable quantitative corrections. In this work we focus
on a particular class of such corrections, namely vertex corrections that can be
parametrized in terms of (off-shell) baryons. In general, baryonic back-reaction effects
onto the quark propagator provide a direct mechanism how the quark condensate
may be influenced by changes in the baryon's wave functions such as the one inflicted
e.g. by the nuclear liquid-gas transition at very small temperatures \cite{Weise:2012yv}.
These back-reaction effects, however, may very well decrease in size for growing
temperatures and it is an open question whether they are still important in the
region of the QCD phase diagram where the putative CEP for the chiral
phase transition is located. In a two-color version of QCD this influence has been
studied in Refs.~\cite{Strodthoff:2011tz,Strodthoff:2013cua} and found to be crucial to an extent that
not only the location but even the very existence of a CEP is affected.
Whether this is also the case in the $SU(3)$ theory is an open question that needs to
be addressed. We regard the study reported in this work as a first step in this direction.

The paper is organized as follows. In Sec.~\ref{sec:truncation} we review our truncation scheme
of the DSEs for the quark and gluon propagators. We explain
the details of the corresponding equation for the quark-gluon vertex and identify
the diagrams that can be parametrized in terms of (non-elementary, i.e. composite)
hadronic degrees of freedom. We specify an approximation scheme for these terms that
is suitable for an exploratory calculation of its effects on dynamical chiral symmetry
breaking. For simplicity we restrict ourselves to the case of QCD ($N_c=3$) with
two degenerate fermion
flavors, $N_f=2$, and note that a generalization to the $N_f=2+1$ case is straightforward
but very expensive in terms of CPU time. In Sec.~\ref{sec:results} we present our
results. We first discuss the effects of the resulting baryon loop on the quark propagator
in the vacuum and at finite temperature but zero chemical potential. We then present an
estimate for the size of the effects that may be expected for the CEP.
We conclude in Sec.~\ref{sec:sum}.

\section{Dyson-Schwinger equations \label{sec:truncation}}

\subsection{DSEs for the propagators \label{sec:truncation1}}

In order to accommodate the notation already for its intended purpose later in this
work, we specify the quark and gluon propagators at finite temperature $T$
and quark chemical potential $\mu_q$ and indicate the limits $T \rightarrow 0$ and $\mu_q \rightarrow 0$
where appropriate. The bare quark propagator is then given by
%\begin{equation}
$S_0^{-1}(p) = i\vect{p}\cdot\vect{\gamma}\,Z_2 + i\tilde\omega_n\gamma_4 \,Z_2 + Z_2 \,m\,,$
%\end{equation}
with wave function renormalisation $Z_2$ and bare quark mass $m$.
The dressed inverse quark propagator $S^{-1}$ and the Landau-gauge gluon propagator
$D_{\mu\nu}$ are given by
\begin{equation}\label{eq:qProp}
\begin{split}
S^{-1}(p) &= i\vect{p}\cdot\vect{\gamma}\,A(p) + i\tilde\omega_n\gamma_4\,C(p) +B(p)\,, \\
D_{\mu\nu}(p) &= P_{\mu\nu}^{T}(p)\frac{Z_{T}(p)}{p^2} + P_{\mu\nu}^{L}(p)\frac{Z_{L}(p)}{p^2}
\end{split}
\end{equation}
with momentum $p=(\omega_n,\vect{p})$. The Matsubara frequencies are $\omega_n=\pi T (2n+1)$ for fermions and
$\omega_n=\pi T \, 2n$ for bosons, and we use the abbreviation $\tilde\omega_n=\omega_n+i\mu_q$.
All dressing functions implicitly depend on temperature and chemical potential.
The projectors $P_{\mu\nu}^{{T},{L}}$ are transverse (${T}$) and longitudinal (${L}$) with respect
to the heat bath and given by
\begin{equation} \label{eq:projTL}
\begin{split}
P_{\mu\nu}^{T} &= \left(1-\delta_{\mu 4}\right)\left(1-\delta_{\nu 4}\right)\left(\delta_{\mu\nu}-\frac{p_\mu p_\nu}{\vect{p}^{\,2}}\right),   \\
P_{\mu\nu}^{L} &= P_{\mu\nu} - P_{\mu\nu}^{T} \,,
\end{split}
\end{equation}
where $P_{\mu\nu} = \delta_{\mu\nu} - p_\mu\, p_\nu/p^2$ is the covariant transverse projector.
The limit of zero chemical potential is straightforward; in the additional zero-temperature limit
the momentum $p$ reduces to its usual $O(4)-$symmetric Euclidean form and the wave functions
of the quark propagator become degenerate, i.e. $A(p)=C(p)$. Furthermore, the
transverse (magnetic) and longitudinal (electric) dressing functions of the gluon
approach the same limit in the vacuum, i.e., $Z_{T}(p)=Z_{L}(p)\equiv Z(p)$. In the medium there exists
also a fourth contribution to the inverse quark propagator, which vanishes in the vacuum and is
proportional to $\tilde\omega_n\gamma_4 \, \vect{p}\cdot\vect{\gamma}$. Due to its negligible contribution
also at higher temperatures and chemical potential we do not consider it throughout this work.

\begin{figure}[t]
\includegraphics[width=0.48\textwidth]{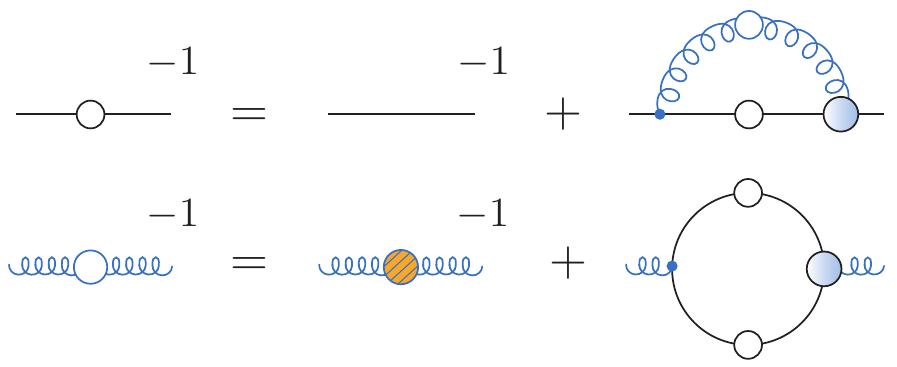}
\caption{The DSE for the quark propagator (top panel) and the
truncated gluon DSE for $N_f=2$ QCD (bottom panel). Large blobs denote dressed
propagators and vertices, and the hatched circle represents the quenched
(lattice) propagator. \label{fig:quarkDSE}\label{fig:apprGluonDSE} }
\end{figure}

The DSE for the quark propagator is shown diagrammatically in Fig.~\ref{fig:quarkDSE}.
The pieces that need to be determined in order to allow for a self-consistent
solution of this equation are the fully dressed gluon propagator and quark-gluon vertex.
Model calculations \cite{Qin:2010nq,Wang:2014yla} often use simple \textit{ans\"atze }for the
gluon propagator that do not take into account the proper temperature and flavor
dependence of the gluon self-energy. We prefer to include these important effects
by taking the Yang-Mills sector of QCD into account and calculating the back-reaction
of the quarks onto the gluon explicitly. This framework has been gradually evolved
from the quenched case, $N_f=0$ \cite{Fischer:2009wc,Fischer:2010fx}, to two-flavor
QCD \cite{Fischer:2011mz,Fischer:2012vc} and recently to $N_f=2+1$ and $N_f=2+1+1$
\cite{Fischer:2012vc,Fischer:2014ata}.

Such an approach has two distinct advantages over
simple modeling. On the one hand it allows us to trace the effects of quark masses and
flavors as exposed in the Columbia plot. On the other hand, it serves to take
into account the effects of chemical potential on the gluon propagator explicitly, thereby
rendering results at finite $\mu_q$ more reliable. Furthermore, since we have
explicit access to all fundamental degrees of freedom of QCD, i.e. quark, gluon and ghost
propagators, we are in a position to determine the
Polyakov loop potential at all values of $T$ and $\mu_q$ and thereby also to study the deconfinement transition. This has
been exploited in Refs.~\cite{Fischer:2012vc,Fischer:2013eca,Fischer:2014ata} for physical
quarks and in Ref.~\cite{Fischer:2014vxa} for a study of the second-order critical surface
of the deconfinement transition of heavy quarks.
These studies are complemented by corresponding ones using the functional renormalization
group, see e.g.~\cite{Christiansen:2014ypa} and references therein.

%%%%%%%%%%%%%%%%%%%%%%%%%%%%%%%%%%%%%%%%%%%%%%%%%%%%%%%%%%%%%%%%%%%%%%%%%%%%
\begin{figure*}[t]
\centering\includegraphics[width=1.0\textwidth]{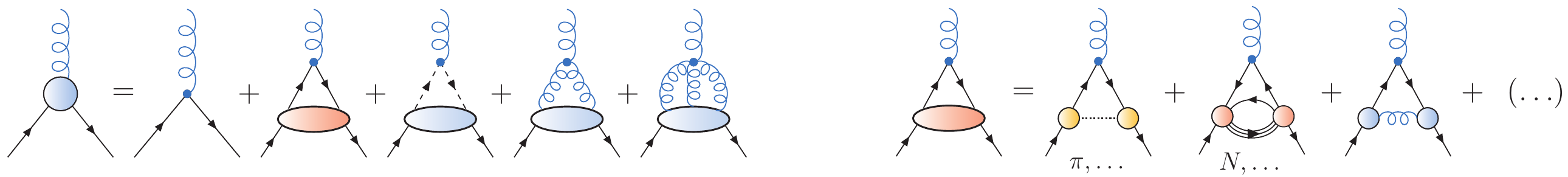}
\caption{The full, untruncated Schwinger-Dyson equation for the quark-gluon
vertex \cite{Marciano:1977su} is shown diagrammatically in the first equation.
The second equation describes the first terms of an expansion in terms of hadronic
and non-hadronic contributions to the quark-antiquark scattering kernel. In both
equations, all internal propagators are fully dressed. Internal dashed lines with
arrows correspond to ghost propagators, curly lines to gluons and full lines to
quark propagators. In the second equation, the dotted line describes mesons and
the triple line baryons.}
\label{fig:Vertexdse}
\end{figure*}
%%%%%%%%%%%%%%%%%%%%%%%%%%%%%%%%%%%%%%%%%%%%%%%%%%%%%%%%%%%%%%%%%%%%%%%%%%%%%

For the gluon DSE, shown in Fig.~\ref{fig:apprGluonDSE}, we use the same setup as described
in detail in Ref.~\cite{Fischer:2012vc}. We use lattice input for the
quenched propagator at different temperatures and determine the temperature and chemical
potential dependent effects of the quark loop explicitly using the quark propagator from
its DSE. We work with two fermion flavors, $N_f=2$, in the isospin limit which allows us
to use one and the same quark DSE for both flavors. The quark loop in the gluon DSE is then
simply multiplied by a factor of two to accommodate for both flavors. The essentials of
this setup are collected in App.~\ref{appUnqGluon}; more details can be found in
\cite{Fischer:2012vc} and shall not be repeated here for brevity.

\subsection{Baryon effects in the quark DSE and quark-gluon vertex \label{sec:truncation2}}

%%%%%%%%%%%%%%%%%%%%%%%%%%%%%%%%%%%%%%%%%%%%%%%%%%%%%%%%%%%%%%%%%%%%%%%%%%%%

\begin{figure}[t]
\centering\includegraphics[width=0.485\textwidth]{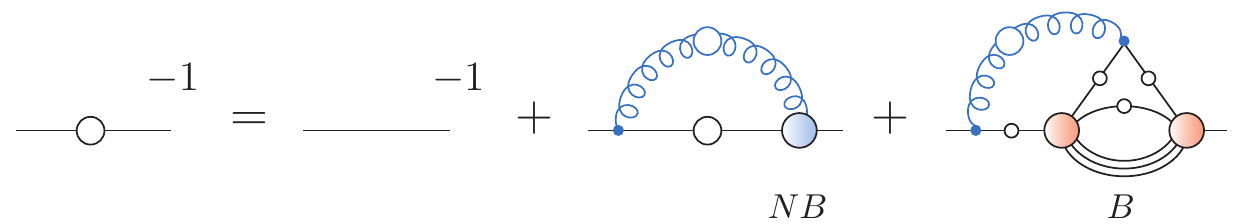}
\caption{DSE for the quark propagator, where the quark-gluon vertex is separated into non-baryon and baryon contributions.}
\label{fig:quark-dse-2}
\end{figure}
\begin{figure*}[t]
\centering\includegraphics[width=0.99\textwidth]{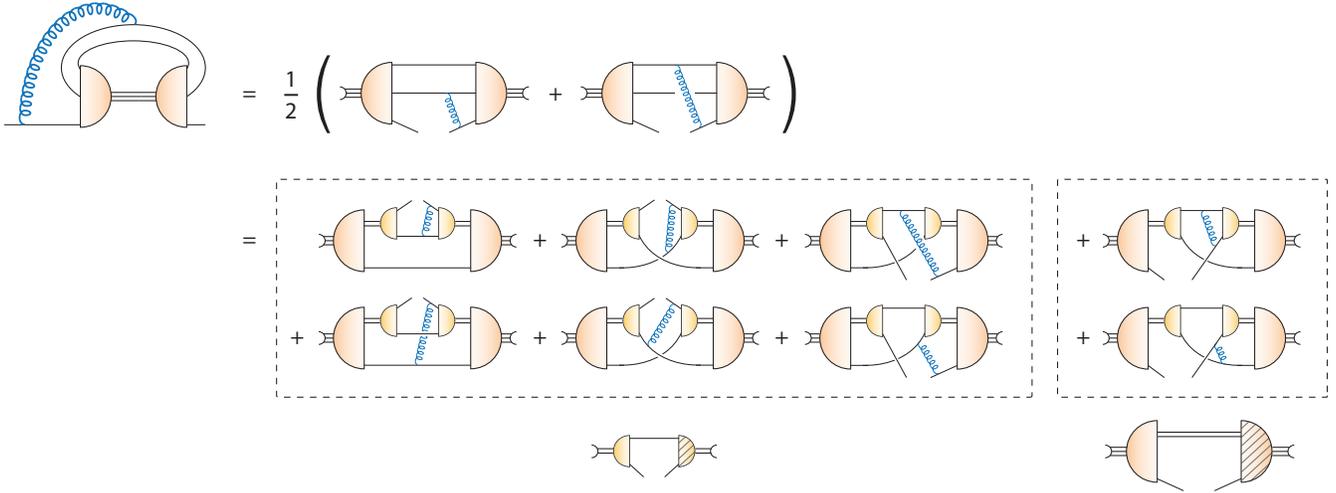}
\caption{Baryonic diagram in the quark DSE with the quark-diquark approximation of the baryon's Faddeev amplitude.}
\label{fig:baryon-diag-approx}
\end{figure*}
\begin{figure*}[t]
\centering\includegraphics[width=0.90\textwidth]{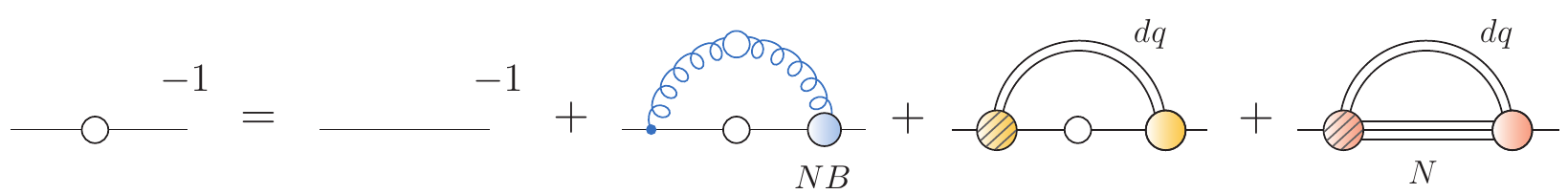}
\caption{Quark DSE with diquark and baryon loop. In these loops the right
vertices (circles) are Bethe-Salpeter amplitudes whereas the left vertices
(hatched circles) are effective ones summing all effects from the
diagrams in Fig.~\ref{fig:baryon-diag-approx}. In the main text we give arguments why these vertices
are well approximated by bare ones.}
\label{fig:quark-dse-3}
\end{figure*}
%%%%%%%%%%%%%%%%%%%%%%%%%%%%%%%%%%%%%%%%%%%%%%%%%%%%%%%%%%%%%%%%%%%%%%%%%%%%

Let us now focus on the Dyson-Schwinger equation for the quark-gluon vertex
and explicate its structure. The full equation is shown in the left part of
Fig.~\ref{fig:Vertexdse}. It contains three one-loop diagrams with a fully
dressed quark (solid), ghost (dashed) and gluon
line (curly) running through the loop and attached to the external gluon by
a corresponding bare vertex. The remaining graph is a gluonic two-loop
diagram with a bare four-gluon vertex. These diagrams contain four- and five-point
Green functions that are 1PI with respect to the external legs in the $t$ channel, i.e., they contain
no contributions from intermediate annihilation of the external quarks into
a single gluon line. The four- and five-point functions can be expanded
in skeleton diagrams with fully dressed internal propagators and primitively
divergent vertices \cite{Marciano:1977su}. For our purposes we concentrate on the
first non-trivial diagram that contains a four-quark amplitude. In its skeleton
expansion, a part of the resulting diagrams can be re-expressed
in terms of Bethe-Salpeter vertices and propagators of mesons as well as
Faddeev-type vertices for baryons.

The result of such an expansion is shown in the second equation in
Fig.~\ref{fig:Vertexdse}. The first diagram corresponds to (off-shell) meson
exchange between the quark lines. It is important to note that this meson is
not introduced as a new elementary field; it is rather a composite object
of a quark and an antiquark that is described (at least on-shell) by its Bethe-Salpeter
equation (BSE).
The first baryon exchange diagram shows up as a two-loop diagram involving the
baryons's Faddeev amplitude. In Ref.~\cite{Fischer:2007ze} the corresponding
diagram has been displayed in the quark-diquark approximation already on the level
of the vertex DSE; below, we will introduce this approximation on the level of
the quark DSE. Finally, we show a representative non-resonant contribution due to
dressed one-gluon exchange. Note that double-counting is trivially avoided in
this combined expansion in elementary and effective degrees of freedom due to
different quantum numbers in the exchange channel.

The computation of the hadronic diagrams in Fig.~\ref{fig:Vertexdse} is rather
involved. The meson-exchange diagram requires the solution of a
coupled system of the DSE for the quark propagator and a
corresponding BSE for the meson Bethe-Salpeter amplitude.
The effect of the pion back-reaction onto the quark in the vacuum has already
been explored to some extent  in the context of pion-cloud contributions to light
mesons and baryons~\cite{Fischer:2007ze,Fischer:2009jm,Sanchis-Alepuz:2014wea}.
Even more complex is the baryon-exchange diagram, which involves the computation
of a Faddeev-type equation for the baryon bound state.
In the present work we are interested in hadronic effects at finite chemical
potential, which will primarily show up in the diagram including baryons because
all elements of this diagram (the quark and baryon propagators as well as the baryon
wave function) depend on chemical potential.
Since the pion-exchange diagram is the hadronic contribution with minimal explicit
dependence upon chemical potential we relegate its explicit study to future work and focus
exclusively on the baryonic diagram.

The resulting quark DSE with an explicit separation into non-baryon and baryon parts
is shown in Fig.~\ref{fig:quark-dse-2}.
The three-loop diagram contained therein is hard to evaluate numerically,
especially at finite temperature and chemical potential. To make this diagram
tractable, we therefore introduce an
additional approximation and convert the three-quark Faddeev amplitude
into a quark-diquark Bethe-Salpeter amplitude for the baryon.
To this end, note that the gluon in the baryon-exchange diagram is attached to the quark on the left,
but since it couples to both quarks symmetrically we can rearrange the diagram as illustrated
by the first equation of Fig.~\ref{fig:baryon-diag-approx}.
The incoming and outgoing baryon lines have to be connected by a baryon propagator
which is indicated by the open circles.

Inserting a separable quark-diquark ansatz for each three-body Faddeev amplitude
leads to the second equality in Fig.~\ref{fig:baryon-diag-approx}. The resulting topologies
can be grouped into two different classes: one where the incoming quark
couples to a diquark amplitude and one where it couples to a quark-diquark amplitude.\footnote{In
principle there are two further diagrams with a closed quark loop where the gluon couples to each of the quarks in the loop. However,
their contributions cancel each other, i.e., the `diquark-gluon vertex' is zero.}
The hatched amplitudes are effective and absorb all the remaining objects in these graphs.
As a consequence, the quark DSE takes the form shown in Fig.~\ref{fig:quark-dse-3}, which contains
a quark-diquark and a baryon-diquark loop. For brevity we will refer to them as `diquark' and `baryon' loops in what follows.
In both diagrams the vertex appearing on the right is a proper Bethe-Salpeter amplitude,
once for a diquark and once for a baryon in quark-diquark approximation. The hatched
vertices on the left carry the same quantum numbers as
their counterparts on the right but represent effective vertices that absorb all effects
appearing in the multi-loop diagrams in Fig.~\ref{fig:baryon-diag-approx}. Thus,
the quark DSE in Fig.~\ref{fig:quark-dse-3}
follows directly from the original equation in Fig.~\ref{fig:quark-dse-2} if the quark-diquark ansatz for the Faddeev amplitude is made.
We will specify its ingredients in Sec.~\ref{sec:truncation3}. For now, note that we work
in the $N_f=2$ theory in the isospin symmetric limit, which leaves us with isospin singlet
scalar and isospin-triplet axial-vector diquarks and a degenerate isospin doublet of nucleons.

For the non-baryonic part of the quark-gluon vertex (denoted by 'NB' in
Fig.~\ref{fig:quark-dse-2}) we employ a construction using
the first term of the Ball-Chiu vertex that satisfies the Abelian Ward-Takahashi
identity~\cite{Ball:1980ay}, multiplied with an infrared-enhanced function of quark and gluon momenta
that accounts for the non-Abelian dressing effects
and the correct ultraviolet running of the vertex. The
explicit expressions are collected in appendix \ref{appYMVertex}. In the following
subsection we complete the discussion of the quark DSE with the last remaining
ingredient, the diquark Bethe-Salpeter amplitudes together with the quark-diquark amplitude for the baryon.

\subsection{Diquark and baryon amplitudes \label{sec:truncation3}}

\begin{figure}[t]
\centering\includegraphics[width=0.45\textwidth]{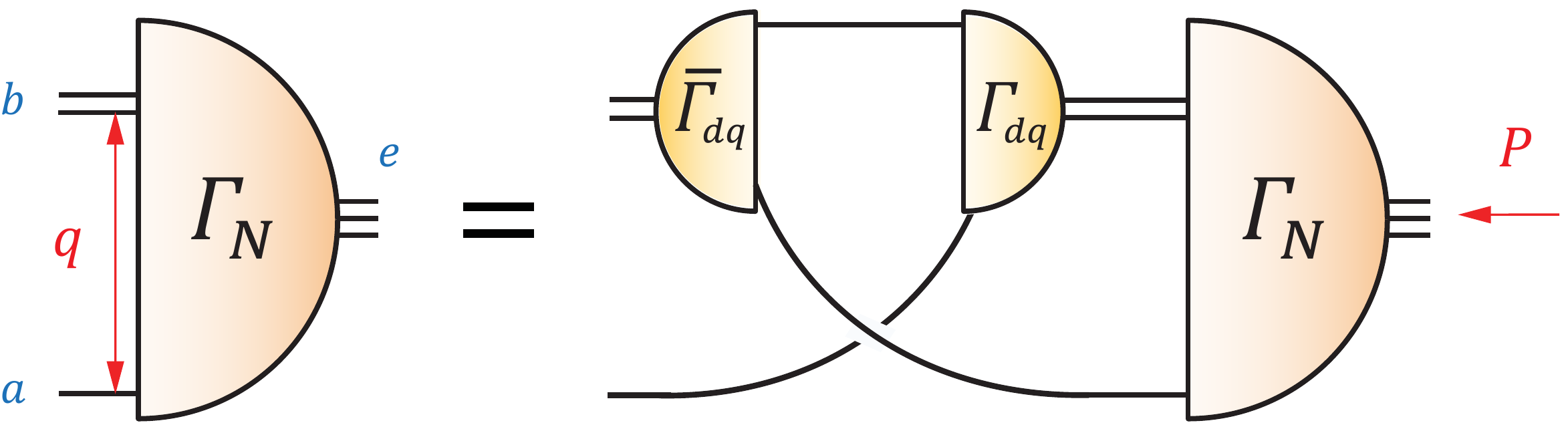}
\caption{The Bethe-Salpeter equation for the baryon in the quark-diquark approximation.}
\label{fig:Faddeev}
\end{figure}

In principle, the baryon is a three-quark state and a comprehensive, full treatment of its
structure should take this explicitly into account. Indeed, the corresponding three-body
Faddeev equation has been solved in
Refs.~\cite{Eichmann:2009qa,Eichmann:2011vu,SanchisAlepuz:2011jn,Sanchis-Alepuz:2014wea,Sanchis-Alepuz:2014sca,Sanchis-Alepuz:2015qra}
and electromagnetic as well as axial form factors have been extracted
\cite{Eichmann:2011vu,Eichmann:2011pv,Alkofer:2014bya}. Owing to the dynamical formation of diquark
correlations inside the nucleon, a potentially satisfying approximation to the three-body
framework is a description in terms of quark and diquark degrees of freedom.
The BSE for such a baryon with quark and diquark constituents is displayed
in Fig.~\ref{fig:Faddeev}. In this approximation, the quark and diquark inside the nucleon
interact via quark exchange and the corresponding diquark amplitude has
to be determined from a separate BSE.

Using a simple model for the underlying quark and diquark propagators and \textit{ans\"atze} for the
diquark amplitudes, baryon properties in the quark-diquark picture
have been determined in many works, see e.g.
\cite{Oettel:1998bk,Oettel:2000jj,Cloet:2008re,Segovia:2014aza} and references therein.
A more fundamental approach is the use of an underlying quark-gluon interaction, from which
all components of such a calculation, the quark propagator in the complex momentum plane,
the diquark amplitude from its BSE, the diquark propagator from its scattering equation,
and the baryon, are
determined consistently without any introduction of further parameters. This has been
performed in \cite{Eichmann:2007nn,Eichmann:2008ef,Nicmorus:2010sd,Eichmann:2011aa}
using a well-established rainbow-ladder interaction kernel for the quark-gluon interaction.
In all these calculations it turned out that a satisfactory description of the ground-state
properties of the nucleon and $\Delta$ baryon can be obtained using scalar and axial-vector
diquarks only. The quark-diquark approximation works well at zero temperature and chemical potential;
below we assume that this is still the case at finite $T$ and $\mu_q$. Whether
that is true remains to be studied in future work.

In fact, if one is only interested in the gross properties of the nucleon even the influence of the
axial-vector diquark may be omitted and both the diquark and the nucleon can be represented
by their leading tensor structure. In this approximation, the diquark and nucleon
Bethe-Salpeter amplitudes are parametrized by
\begin{equation}\label{BSAs}
\begin{split}
\Gamma_{dq}(q,P) &= f_{dq}(q^2) \,\gamma^5 C \otimes \frac{\epsilon_{ABE}}{\sqrt{2}}
\otimes s^0_{ab}\,,\\
\Gamma_N(q,P) &= f_N(q^2) \,\Lambda_+(P) \otimes \frac{\delta_{AB}}{\sqrt{3}} \otimes t^0_{ae}\,.
\end{split}
\end{equation}
Here, $q,P$ are the relative and total momentum of the bound states, $C=\gamma_4\gamma_2$ is the charge-conjugation
matrix, and $\Lambda_+$ the projection operator onto positive-energy states
(which we omit in the baryon loop diagram because its purpose is already served by the nucleon propagator).
We use normalized color wave functions with capital subscripts and
normalized flavor wave functions with small subscripts;
$s^0 = \frac{1}{\sqrt{2}}(u d^\dagger - d u^\dagger) = \frac{1}{\sqrt{2}}i\sigma_2$ with Pauli
matrices $\sigma_i$ and $t^0 = (u u^\dagger + d d^\dagger) = \mathbbm{1}$.
The solutions for the diquark and nucleon amplitudes determined in the rainbow-ladder
framework of Refs.~\cite{Eichmann:2008ef,Nicmorus:2010sd,Eichmann:2011aa} are well parametrized by
\begin{equation} \label{bar}
\begin{split}
f_{{dq}}(q^2) &= N_{dq} \left(e^{-\alpha_{dq} \cdot x} + \frac{\beta_{dq}}{1+x}\right)\,,\\
f_N (q^2) &= N_{N} \left(e^{-\alpha_{N} \cdot x} +	\frac{\beta_{N}}{(1+x)^3}\right)
\end{split}
\end{equation}
with $x=q^2/\Lambda^2$ and the scale $\Lambda=0.7$ GeV. The normalization factors
are obtained from normalizing the corresponding full Bethe-Salpeter amplitudes and are given by $N_{dq}=15.6$
and $N_{N}=28.4$. The parameters are $\alpha_{dq}=0.85$ and $\alpha_{N}=1.0$ for the exponentials and $\beta_{dq}=0.02$ and $\beta_{N}=0.03$ for the UV behavior.

The remaining question concerns the effective amplitudes in Fig.~\ref{fig:quark-dse-3} which we did not specify yet.
For those we resort to a simple approximation:
we take them as `bare', i.e.,
we use Eq.~\eqref{BSAs} with $f_{dq}(q^2) = f_N(q^2)=1$.
This is analogous to the treatment of the pion loop in the quark DSE in Refs.~\cite{Fischer:2007ze,Fischer:2009jm,Sanchis-Alepuz:2014wea}.
It can be motivated by estimating the overall strength of the baryon diagram in Figs.~(\ref{fig:quark-dse-2}--\ref{fig:baryon-diag-approx}) from its contribution to the quark condensate.
When connecting the quark lines with the scalar $q\conjg{q}$ vertex (calculated from its inhomogeneous
BSE), the resulting vacuum bubble $\mathcal{C}$ is proportional to the integrated (off-shell) scalar form factor of the nucleon:
              \begin{equation}
                  \mathcal{C} = \frac{2m_N}{3}\!\int \!\!\frac{d^4P}{(2\pi)^4} \, \frac{g_S(P^2,Q^2=0)}{P^2+m_N^2}\,.
              \end{equation}
This can be seen by inserting the covariant Faddeev equation for the three-body amplitude in the first line of Fig.~\ref{fig:baryon-diag-approx};
the resulting quantity is what appears in baryon form factor diagrams such as in Ref.~\cite{Eichmann:2011vu}.
The on-shell value of the scalar form factor at $P^2=-m_N^2$ is determined by the nucleon sigma term via the Feynman-Hellmann theorem:
              \begin{equation}
                 \sigma_N = m_q \,g_S(-m_N^2,0) = m_q\,\frac{dm_N}{dm_q} \approx m_\pi^2\,\frac{dm_N}{dm_\pi^2}\,.
              \end{equation}
The magnitude of $\mathcal{C}$ obtained with the experimental value $\sigma_N = 45$ MeV, together
with an integral cutoff at $m_N=0.94$ GeV, is similar to the value obtained from numerically tracing the sum of the
diquark and baryon diagrams in Fig.~\ref{fig:quark-dse-3} with a scalar vertex, however  with only \textit{one} amplitude dressed in each case.\footnote{We
also calculated the vacuum bubble from the baryon diagram in Fig.~\ref{fig:quark-dse-2} directly using a reasonable off-shell ansatz for the three-quark Faddeev amplitude instead of a cutoff;
the result is in the same ballpark.}
Dressing both would overestimate the strength by far due to the normalization factors in Eq.~\eqref{bar}.
Since we are only interested in the gross effects of baryons on the phase diagram, we therefore view this
as a justified approximation.

\subsection{Quark DSE including hadronic loops \label{sec:truncation4}}

Putting everything together,
we will now give explicit expressions for the diagrams in the quark and gluon DSEs
including the hadronic back-reaction diagrams. As will become clear below, the hadronic effects
on the quark propagator enter on the level of ten-percent
corrections. From the diagrammatic form of the quark-gluon vertex DSE this is exactly the order of magnitude as expected, since
the corresponding diagrams are suppressed by a factor $1/N_c^2$. In the Yang-Mills
sector of QCD the total quark effects are on the level of a $1/N_c$ correction.
Therefore, hadronic contributions to the quark-loop diagram in the gluon DSE only
contribute at $1/N_c^3$ and it is well justified to neglect those in
a first exploratory calculation. Hence we will use the same truncation for the
gluon DSE as in previous works \cite{Fischer:2011mz,Fischer:2012vc,Fischer:2014ata}.

In the quark DSE we take into account the three diagrams in Fig.~\ref{fig:quark-dse-3}.
If we denote the quark dressing functions in Eq.~\eqref{eq:qProp} collectively by $H(p) = A(p), B(p), C(p)$
and abbreviate the gluon, diquark and baryon-loop contributions to the quark self-energy by $\Sigma_H^{glue}$, $\Sigma_H^{dq}$ and
$\Sigma_H^{ba}$, the resulting equations read
\begin{equation}
  H(p) = Z_2\,\lambda_H + \Sigma_{H}^{glue} + \Sigma_{H}^{dq} + \Sigma_{H}^{ba} \,,
\end{equation}
where $\lambda_B=m_0$ is the bare current-quark mass, $\lambda_A=\lambda_C=1$, and $Z_2$ is the quark
wave-function renormalization constant.
The gluon-dressing loop $\Sigma_{H}^{glue}$ contains the
unquenched, temperature- and chemical-potential dependent gluon propagator together with
a model for the quark-gluon vertex~\cite{Fischer:2011mz,Fischer:2012vc,Fischer:2014ata};
the explicit formulas are relegated to App.~\ref{appYMVertex}.
The self-energy contributions from the diquark and baryon loop
are given by
   \begin{equation}\label{dq-ba-self-energy}
   \begin{split}
     \!\Sigma_{H}^{dq}(p) &= \frac{1}{2}   \is_q\,
                                         \frac{f_{{dq}}(\frac{q-p}{2})\,D_{dq}(q+p)}{\vect{q}^2 A^2(q) + \tilde{\omega}_q^2 C^2(q) + B^2(q)} \,K_{H}^{dq}, \\
     \!\Sigma_{H}^{ba}(p) &= \frac{1}{3}   \is_q\,
                                       \frac{f_{{N}}(\frac{q}{2}-p)\,D_{dq}(q-p) }{\vect{q}^2  + (\omega_q + 3i\mu_q)^2 +  m_N^2} \,K_{H}^{ba}
   \end{split}
   \end{equation}
   with
   \begin{alignat}{6}
     &&  K_A^{dq} &= \frac{\vect{p}\cdot\vect{q}}{\vect{p}^2}\,A(q)\,,  & \qquad
         K_A^{ba} &= \frac{\vect{p}\cdot\vect{q}}{\vect{p}^2}\,, \nonumber \\
     &&  K_C^{dq} &= \frac{\tilde{\omega}_q}{\tilde{\omega}_p}\,C(q)\,,  & \qquad
         K_C^{ba} &= \frac{\omega_q+3i\mu_q}{\tilde{\omega}_p}\,,  \\
     &&  K_B^{dq} &= -B(q)\,, & \qquad
         K_B^{ba} &= -m_N\,. \nonumber
   \end{alignat}
We have already carried out all color and flavor traces. At finite temperature and
chemical potential the arguments $p$, $q$ serve as abbreviations for $p=(\omega_n,\vect{p})$, $q=(\omega_m,\vect{q})$.
The (fermionic) Matsubara frequencies are given by $\omega_n = \pi T (2n+1)$ and we write
$\tilde{\omega} = \omega + i\mu_q$ with quark chemical potential $\mu_q$.
The Matsubara sum as well as the
integration over the loop three-momentum $\vect{q}$ is abbreviated by
$\is_{\;\;q} = T\sum_{n_q} \int\frac{d^3q}{(2\pi)^3}$, and the diquark propagator $D_{dq}$ is given by
\beq
D_{dq}(q \pm p) = \frac{1}{(\vect{q} \pm \vect{p})^2+(\omega_q \pm \omega_p+2i\mu_q)^2+m_{{dq}}^2}\,.
\eeq

In these expressions we take into account the lowest-lying $J^P = \nicefrac{1}{2}^+$ baryon multiplet
for the two-flavor case, i.e., the nucleon,
in the approximation with scalar diquarks only. In principle other baryons
may also contribute but since they are suppressed by powers of $m_N^2/m_B^2$
with respect to the nucleon their influence is certainly subleading.
There is, however, one exception:
the parity partner of the nucleon becomes (approximately) mass-degenerate
once chiral symmetry is restored, i.e. in the high temperature/density phase. Performing
the Dirac traces of the corresponding loops, it turns out that contributions from
mass-degenerate multiplets of parity partners cancel each other in $\Sigma_B^{ba}$, while
they add up in the other two contributions to the quark self-energy. We take this
effect qualitatively into account by multiplying the right-hand side of $\Sigma^{ba}_B$
with an additional factor $M(T,\mu_q)/M(0,0)$ evaluated at zero momentum and lowest
Matsubara frequency. Here $M=B/A$ defines the renormalization-point
independent quark mass function in the medium. This factor has no effect in the vacuum but mimics the
cancellation of multiplets of parity partners in the chirally restored phase,
where $M(T,\mu_q)$ becomes small. A corresponding factor of $2-M(T,\mu_q)/M(0,0)$
is added to $\Sigma_A^{ba}$ and $\Sigma_C^{ba}$. Note that the diquark loop, which
has been derived from the original baryon three-body diagram via the quark-diquark picture of baryons, contains
diquarks only; thus there are no
damping/enhancement factors in $\Sigma_{A,B,C}^{dq}$.

The remaining unknowns in Eq.~\eqref{dq-ba-self-energy} are the temperature and chemical-potential
dependence of the diquark and baryon masses and Bethe-Salpeter amplitudes. Ideally these need
to be determined consistently from their BSEs evaluated at
finite $T$ and $\mu_c$. This formidable numerical task is yet to be
performed and relegated to future work. Here, in this exploratory work, we resort to the
vacuum expressions for the diquark and nucleon amplitudes as given in Eq.~\eqref{bar},
evaluated at four-momenta that include temperature effects in the form
of Matsubara frequencies and the results for the masses from the corresponding bound state calculations
$m_N=0.938$ GeV and $m_{dq}=0.810$ GeV.
Certainly this can only be a first approximation on a qualitative
level. In order to gauge the quantitative effects of including potential changes of the
baryon and diquark masses and wave functions with chemical potential, we will introduce
and discuss additional dependencies on $\mu_q$ in section \ref{sec:results3}.

\section{Results \label{sec:results}}

\begin{figure*}[t]
\centering
\includegraphics[width=0.99\textwidth]{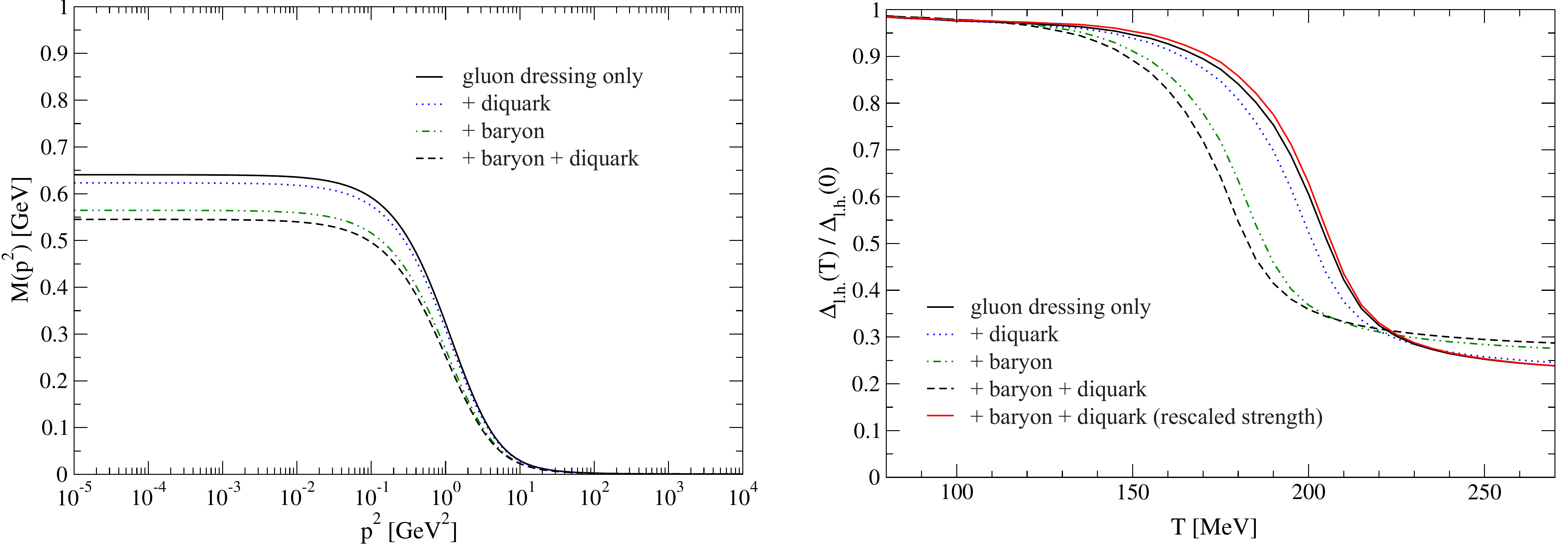}
\caption{\textit{Left:} quark mass function with and without diquark and baryon loops included.
\textit{Right:} regularized and normalized condensate as a function of temperature with and
without diquark and baryon loops. In addition, we display the result
with rescaled strength in the quark-gluon interaction, see main text for details.}
\label{fig:mu0}
\end{figure*}

\subsection{Vacuum \label{sec:results1}}

Before we consider baryon effects at finite temperature and chemical potential, we
first study the impact of the different loops on the strength of dynamical chiral
symmetry breaking in the vacuum. Both diquark and baryon loops
originate from diagrams in the quark-gluon vertex DSE that contain additional
quark loops. In general these are chirally restoring, as has been discussed in
Refs.~\cite{Fischer:2005en,Fischer:2007ze}. On the lattice, the unquenched quark
propagator has indeed a smaller mass function than the quenched one \cite{Kamleh:2007ud}.
While this behavior is expected for the total sum of all unquenching contributions
to the quark propagator, it is not necessarily true for each individual diagram
such as those investigated here. As a first exercise we therefore study the
sign and magnitude of the individual effects of each contribution onto the quark mass function.
For the calculation we use two dynamical quark flavors in the gluon DSE, $N_f=2$,
a renormalized bare quark mass $m(\mu^2)=0.8$ MeV at the renormalization point
$\mu= 80$ GeV, and an interaction strength parameter $d_1=8.05$ GeV$^2$ in the
quark-gluon vertex (cf. App.~\ref{appYMVertex}). These values have been taken over
from the $N_f=2+1$ theory (set $A_{2+1}$ in Ref.~\cite{Fischer:2014ata}, matching
corresponding lattice results) without further adaption for reasons of simplicity.

Our results are displayed in the left diagram of Fig.~\ref{fig:mu0}. It shows the quark
mass function calculated with only the usual gluon-dressing loop, compared to results
including the diquark and baryon loops individually and in combination. Indeed, all
additional contributions are chirally restoring, with a larger effect coming from the
baryon loop. The total contribution of both loops is $\sim 15\%$; they reduce the
quark mass function at zero momentum from $M(0)=640$ MeV to $M(0)=545$ MeV, whereas
the impact is immaterial in the large momentum regime.\footnote{Note
that the considerable size of the generated quark mass is a direct effect of performing a
$N_f=2$-calculation while working with scales adapted to the $N_f=2+1$ theory.
An additional back-coupling of the strange quark would reduce the strength of the gluon
propagator and decrease the quark mass considerably.} 
In total, we find that the
effect due to baryons is rather large. Lattice QCD finds total
unquenching effects in the quark mass function of less then $20~\%$~\cite{Kamleh:2007ud};
thus our baryonic effects leave almost no room for other unquenching corrections like e.g.
meson back-coupling effects. This may be attributed to the
comparably simple approximation of the baryon wave functions used in this work.
On the other hand, one of the goals herein is to gauge the systematic effects
of such contributions onto the QCD phase diagram. In such a study it seems better to
over- than to underestimate the induced systematic corrections.

\subsection{Finite temperature \label{sec:results2}}

Next we assess the effects of the diquark and baryon loops on the quark condensate
evaluated at finite $T$. We use a regularized expression for the condensate,
\begin{equation}
\Delta_{l,h} = \langle\bar\psi\psi\rangle_l - \frac{m_l}{m_h}\langle\bar\psi\psi\rangle_h\,,
\label{eq:cond_renorm}
\end{equation}
which eliminates the divergences appearing for non-zero bare quark masses.
The definition of $\langle\bar\psi\psi\rangle$ is given in Eq.~\eqref{eq:condensate}.
For the heavy quark mass we choose $m_h(80 \,\mbox{GeV})= 100$ MeV. In order to evaluate
the corresponding condensate $\langle\bar\psi\psi\rangle_h$ in the $N_f=2$ theory we
would need to solve the complete coupled system of DSEs, Fig.~\ref{fig:quarkDSE}, a
second time for each temperature and chemical potential. However, the sole purpose of
$\langle\bar\psi\psi\rangle_h$ is regularization. Thus it turns out to be sufficient
to evaluate this quantity from the quark DSE with modified quark mass $m_l \rightarrow m_h$
in the bare quark propagator $S_0^{-1}$, but keeping the gluon and the quark-gluon
vertex (including baryonic loops) from the light-quark calculation.
We have explicitly checked that this procedure is a good approximation for some
selected values of temperature and chemical potential and then adopted it throughout
the phase diagram. The transition temperatures for the chiral crossover
are extracted from the maximum of the chiral susceptibility.

Our results for $\mu_q=0$ are displayed in the right diagram of Fig.~\ref{fig:mu0}. Compared to the
calculation without diquark and baryon loops, the additional loops reduce the strength
of dynamical chiral symmetry breaking and the transition temperature for the chiral
crossover reduces correspondingly. It turns out that this sizable impact of the baryonic
contributions on the chiral transition can be almost completely reabsorbed into the vertex
truncation by rescaling the strength of the 'NB'-part of the vertex. To this end we modify
the parameter $d_1$ (cf.~Eq.~(\ref{eq:vertex2})) such that the critical temperature does
not change at $\mu_q=0$ upon taking baryon loops explicitly into account. This amounts
to $d_1=8.05$ GeV$^2 \rightarrow$ $d_1=8.94$ GeV$^2$. The resulting
condensate (the red solid curve in the plot), where baryon and diquark effects are included,
recovers to very good accuracy the original shape of the condensate.

This observation is important for our general strategy. In
Ref.~\cite{Fischer:2014ata} the lattice data for the condensate of the $N_f=2+1$
theory have been reproduced point-wise in a formulation using the gluon-dressing loop only, without
making the baryonic degrees of freedom explicit. Here, for $N_f=2$, we observe
that we can reproduce a similar functional dependence of $\Delta_{l,h}(T)$ using
explicit baryonic degrees of freedom and a rescaled version of the quark-gluon
interaction. We regard this as a strong indication that the same property holds in
the $N_f=2+1$ theory.
In the following, we therefore use the $N_f=2$ theory with
rescaled interaction strength parameter $d_1$ as a template to study the baryonic effects
at finite $\mu_q$.

While at zero chemical
potential all effects can be absorbed into $d_1$, this is not {\it a priori} clear for the
finite chemical potential case because the diquark and baryon loops contain
a much stronger explicit dependence on $\mu_q$ than the gluon dressing loop, as
discussed above. In the next section we will explore the consequences of these additional
contributions for the location of the critical end point (CEP).

\begin{figure}[t]
\centering
\includegraphics[width=0.48\textwidth]{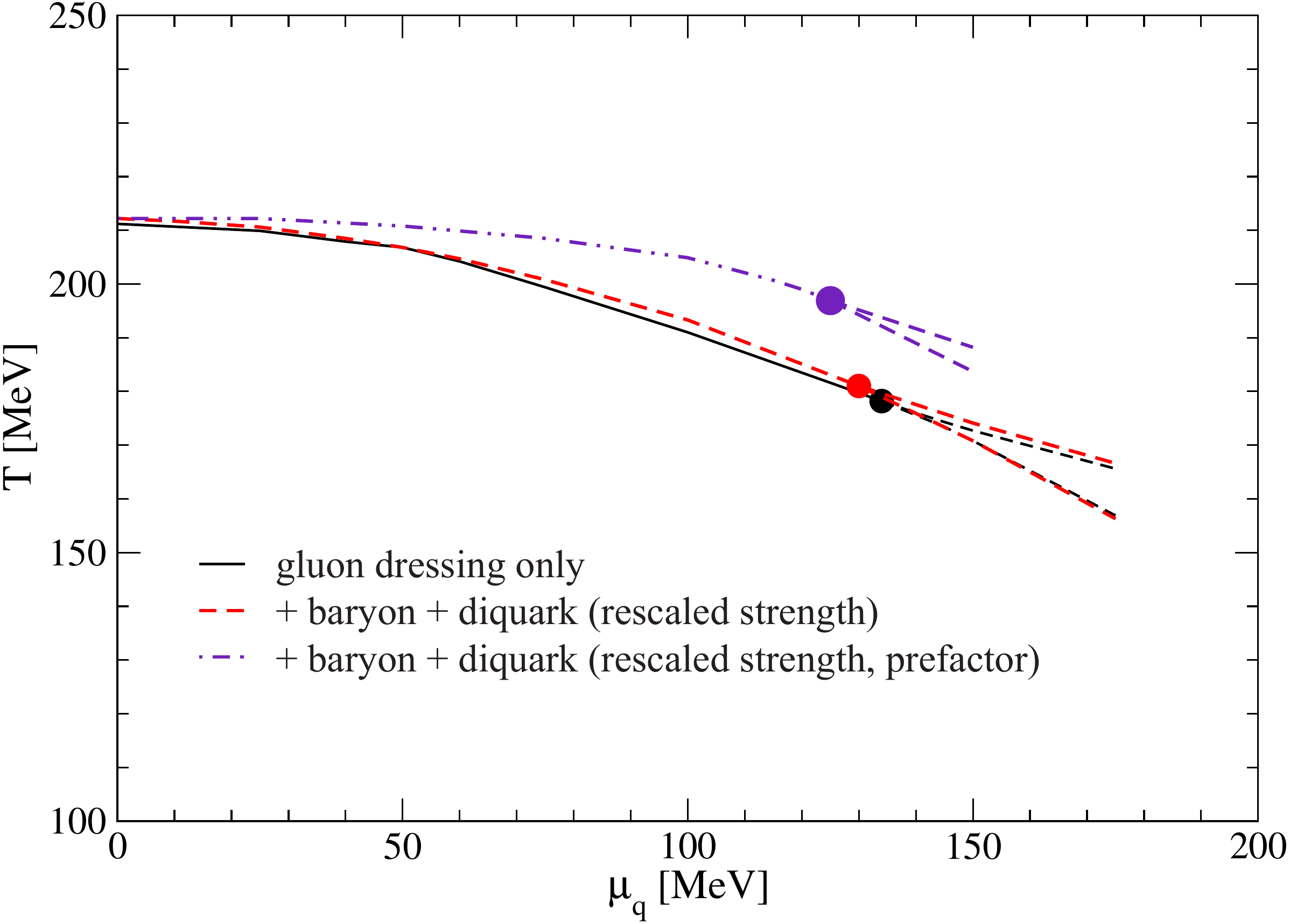}
\caption{Comparison of the phase diagram for $N_f$=2 including different types of selfenergy contributions.}
\label{fig:pd}
\end{figure}

\subsection{Finite temperature and chemical potential \label{sec:results3}}
In the Dyson-Schwinger approach to the QCD phase diagram the introduction of (real) quark chemical potential
is straightforward; cf.~Eq.~\eqref{eq:qProp}, where the dressing functions of the quark propagator become complex.
The impact of chemical potential is apparent in the quark propagator but it also affects the gluon (cf. App.~\ref{appUnqGluon})
as well as the quark-gluon vertex (cf. App.~\ref{appYMVertex}) due to the explicit unquenching procedure.
In this section we investigate how this nontrivial influence is modified by the baryon and
diquark loops. We use the maximum of the chiral susceptibility, Eq.~\eqref{eq:TCchiralsusz}, as the definition of the
(pseudo-)\,critical temperature.

In Fig.~\ref{fig:pd} we show the results in the T-$\mu_q$ plane. The solid (black) curve
is the result for the unquenched system with gluon-dressing loop and no baryonic effects.
For small values of the chemical potential $\mu_q=0$ the transition is a cross-over up
to the filled circle, which indicates the 2nd order critical end point (CEP)
at the critical value $\mu_q = \mu^c_q$. The two dashed lines emerging from the CEP mark
the first order spinodal region for $\mu_q > \mu^c_q$. In this case we find a critical endpoint at
\beq
(T^c,\mu^c_q)=(177, 134) \,\mbox{MeV}
\eeq
for the two-flavor theory $N_f=2$.
Comparison with the dashed (red) curve, which includes explicit baryonic effects with rescaled vertex strength,
leads us to the following observations:
\begin{itemize}
 \item a critical endpoint still exists;
 \item the chiral phase transition lines are almost on top of each other;
 \item the critical endpoint is shifted by less than 5 MeV to smaller chemical potential.
\end{itemize}

The first observation is important because in QC$_2$D a disappearance of the critical
endpoint was observed after introducing the two-color equivalent of baryonic
effects~\cite{Strodthoff:2011tz,Strodthoff:2013cua}, see also
\cite{Cotter:2012mb,Boz:2013rca,Boz:2015ppa}.
While a negligible influence of baryonic degrees of freedom on the transition line is generally expected at low chemical potential,
in our case such a behavior also persists for higher $\mu_q$ and remains
true for the critical endpoint. This also implies that our original truncation with the unquenched gluon-dressing
loop can implicitly absorb baryonic effects at finite
chemical potential, at least those that are captured by our simple approximation.

Potentially important effects beyond our current scheme are additional dependencies of the baryon's
mass and wave function on the chemical potential. Unfortunately not much is know in this respect. In Ref.~\cite{Wang:2013wk} the authors investigate the thermal
properties of baryons at $\mu_q=0$ and find that the amplitudes are almost independent of $T$,
whereas the masses rapidly increase around the (pseudo-)\,critical temperature. As a result, baryonic
contributions would decrease in importance in a region close to the transition line. Such temperature-dependent
effects would presumably have no impact on our results since they can
be reabsorbed in the strength of the 'NB'-part of the vertex.

In the absence of explicit knowledge, we gauge the impact of modifications of the baryon wave function
with chemical potential by multiplying the baryon loop with a function
\beq\label{fkappa}
f_\kappa(\mu_q) = 1-\frac{\mu_q/\Lambda_\kappa}{1+a_\kappa(\mu_q/\Lambda_\kappa)+b_\kappa(\mu_q/\Lambda_\kappa)^2}\,,
\eeq
where we make use of recent evaluations
of the curvature of the chiral transition line on the lattice \cite{Bonati:2015bha,Bellwied:2015rza,Cea:2015cya}.
This curvature can be parametrized in terms of a quantity $\kappa$,
\beq
\frac{T(\mu_B)}{T(0)} = 1 - \kappa \left(\frac{\mu_B}{T(\mu_B)}\right)^2,
\eeq
that characterizes the lowest order in a Taylor expansion in the baryonic chemical potential $\mu_B$. While
recent lattice values for $\kappa$ range between $0.0135 \dots 0.020$, our value is somewhat larger.
Neglecting $1/N_f$ corrections, we adopt the $N_f=2+1$ value $\kappa = 0.0149$ from~\cite{Bellwied:2015rza}
for our $N_f=2$-calculation and match the coefficients $\Lambda_\kappa,a_\kappa,b_\kappa$ in Eq.~(\ref{fkappa})
such that we reproduce the lattice curvature in a region where the lattice can well
be trusted. This is possible for
$\Lambda_\kappa=0.714$ GeV, $a_\kappa=-10.3$ and $b_\kappa=36$. The function $f_\kappa$ then has a minimum
at $\mu_q \approx 120$ MeV. For larger chemical potential we use $f_\kappa(\mu_q) = f_\kappa(\mu_q=120 \,\mbox{MeV})$
to make the function monotonic.
The resulting phase diagram is shown as the dash-dotted (indigo) curve in Fig.~\ref{fig:pd}, with a new location
of the critical end point at
\beq
(T^c,\mu^c_q)=(197, 125) \,\mbox{MeV}\,.
\eeq
Due to the smaller curvature at low chemical potential, the CEP shifts by $\sim 10\%$ towards larger
temperature and $\sim 5\%$ towards smaller chemical potential. The ratio $\mu_B^c/T^c$ changes
accordingly from $\mu_B^c/T^c = 2.3$ to $\mu_B^c/T^c = 1.9$. These changes are by no means dramatic but
quantitatively significant. However, this comes at the expense of a modification of the strength of the
baryon loop by more than $50\%$ due to the additional function $f_\kappa$. Whether such a variation of the
baryon wave function and masses with chemical potential is realistic or not needs to be investigated in
the future. It is also by no means clear whether baryonic effects are the only possible source
of a smaller curvature in the present framework. Here we only demonstrated that it is possible in
principle that baryonic corrections can induce such an effect.

Qualitatively, it is always the case that the CEP shifts towards larger temperatures and smaller chemical
potential if $f_\kappa(\mu_q)$ is smaller than one for $\mu_q > 0$. The opposite effect can be
obtained if $f_\kappa(\mu_q)$ is chosen to be larger than one: the CEP then shifts towards smaller
temperatures and larger chemical potential. Apart from arguments by comparison with the lattice
we see no physical reason a priori why baryon effects should have one effect or the other. Again,
this needs to be studied in a more advanced framework.

\section{Summary and conclusions \label{sec:sum}}
In this exploratory study we extended our existing truncation of the coupled system for the quark and
gluon Dyson-Schwinger equations to take explicit baryonic degrees of freedom into account. This was
achieved by considering a specific class of diagrams in the Dyson-Schwinger equation for the quark-gluon
vertex where genuine hadronic contributions can be identified. Upon introducing baryons through the
quark-diquark picture, the baryon diagrams enter as a baryon-diquark loop
and a quark-diquark loop in the quark Dyson-Schwinger equation in addition to the
gluon-dressing loop. In the $N_f=2$ calculation performed herein we employ the vacuum amplitudes and masses for
the nucleon and the (scalar) diquark, but we take into account cancellation effects due to the degeneration
of the chiral partner of the nucleon through a factor that couples the baryon-diquark loop to the
chiral dynamics of the system. With this setup we performed a calculation of the QCD phase diagram and find
that the inclusion of baryon degrees of freedom changes the location of the critical endpoint only by a
few MeV in $T$ and $\mu_q$.

More drastic effects are possible once a dependence of the baryon masses and wave functions
on chemical potential are taken into account. We estimated these using a parametrization that reproduces
the lattice transition line at small chemical potential. As a result we find a shift of the CEP in the
$5\dots 10\%$ range which drives the ratio $\mu_B^c/T^c$ slightly below 2. We expect that these
results obtained in the two-flavor theory still hold qualitatively for the $N_f=2+1$ case. This will be
explored in future work, where we also strive to determine the chemical potential dependence of the baryon
and diquark masses and wave functions explicitly.

\section*{ACKNOWLEDGEMENTS}
We thank Bernd-Jochen Schaefer and Lorenz von Smekal for fruitful discussions.
Furthermore we thank Christian~H.~Lang for contributions in the early stages of
this project. This work has been supported by the Helmholtz International Center
for FAIR within the LOEWE program of the State of Hesse and by the German Science
Foundation DFG under project number TR-16.

\vspace*{5mm}
\begin{appendix}

\section{Gluon contribution to quark self-energy} \label{appYMVertex}

To make the paper self-contained we collect in this appendix the ingredients
of the non-baryonic (NB) part of the quark DSE, i.e., the gluon-dressing loop in Fig.~\ref{fig:quark-dse-3}.
Our ansatz for the quark-gluon vertex that appears therein is given by
\begin{equation} \label{eq:vertex1}
\Gamma_\mu(p,q;k) = \gamma_\mu\,(\delta_{\mu i}\,\Gamma_S + \delta_{\mu 4}\,\Gamma_4)\,\Gamma(k^2)  \,,
\end{equation}
where
\begin{equation} \label{eq:vertex2}
\begin{split}
  \Gamma_S &= \frac{A(p)+A(q)}{2}\,, \qquad \Gamma_4 =\frac{C(p)+C(q)}{2}\,, \\
 \Gamma(k^2) &= \frac{d_1}{d_2+k^2} + \frac{x}{x+1}
\left[\frac{\beta_0 \,\alpha}{4\pi}\,\ln(1+x)\right]^{2\delta}
\end{split}
\end{equation}
with $x=k^2/\Lambda^2$.
Here, $p=(\omega_p,\vect{p})$ and $q=(\omega_q,\vect{q})$ are the fermion momenta and
$k=(\omega_k,\vect{k})$ is the gluon momentum. The scales $d_2 = 0.5 \,\mbox{GeV}^2$
and $\Lambda = 1.4\,\mbox{GeV}$ and the coupling $\alpha=0.3$ are adapted to match the
corresponding scales of the quenched lattice gluon propagator that we use in the
gluon DSE. Whereas $d_2$ and $\Lambda$ control the renormalization-group running
of the vertex function from the large- into the low-momentum region, $d_1$ controls
the strength of the quark-gluon interaction at small momenta and therefore the
amount of dynamical chiral symmetry breaking in the hadronic phase. Its value is
discussed in the main text.
The ultraviolet momentum region is governed by the anomalous dimension of the vertex $\delta=-9\frac{Nc}{44N_c - 8N_f}$
and $\beta_0=\frac{11N_c-2N_f}{3}$.

The explicit expressions
for the gluonic parts of the self energy are:
\begin{align}
\Sigma_A^{glue} &= Z_2 \,C_F \,g^2 \is_q\frac{\Gamma(k^2)}{D(q)}\,\frac{ A(q)K_{AA}+ C(q)K_{AC}}{\vect{p}^2}\,, \nonumber \\
\Sigma_B^{glue} &= Z_2 \,C_F \,g^2 \is_q\frac{\Gamma(k^2)}{D(q)}\,B(q)\,K_{BB}\,, \\
\Sigma_C^{glue} &= Z_2 \,C_F \,g^2 \is_q\frac{\Gamma(k^2)}{D(q)}\,\frac{A(q)K_{CA} + C(q)K_{CC}}{\tilde{\omega}_p}\,, \nonumber
\end{align}
where $k=p-q$, $C_F=\frac{4}{3}$ is the Casimir operator, $Z_2$ is the wave-function renormalization constant,
and $\Gamma$ has been defined above. The Matsubara sum as well as the
integration over the loop three-momentum $\vect{q}$ is represented by
$\is_{\;\;q} = T\sum_{n_q} \int\frac{d^3q}{(2\pi)^3}$. The denominator of the quark propagator
is given by $D(q) = \vect{q}^2 A^2(q) + \tilde{\omega}^2_q \, C^2(q) + B^2(q)$ and the kernels
$K$ read
\begin{equation}
\begin{split}
K_{AA} &= \Gamma_S\,\bigg[\frac{Z_L}{k^2}\,\frac{\omega_k^2}{k^2}\left(\vect{p}\cdot\vect{q} - 2\,\frac{\vect{p}\cdot\vect{k}\;\vect{q}\cdot\vect{k}}{\vect{k}^2}\right)  \\
       & \qquad + 2\,\frac{Z_T}{k^2}\frac{\vect{p}\cdot\vect{k}\;\vect{q}\cdot\vect{k}}{\vect{k}^2} \bigg]  + \Gamma_4\, \frac{Z_L}{k^2}\frac{\vect{k}^2}{k^2}\,\vect{p}\cdot\vect{q}\,,  \\
K_{AC} &= (\Gamma_S+\Gamma_4)\,\frac{Z_L}{k^2}\,\frac{\vect{p}\cdot\vect{k}}{k^2}\,\tilde{\omega}_q\,\omega_k\,, \\
K_{BB} &= \Gamma_S\left(2\frac{Z_T}{k^2} + \frac{Z_L}{k^2}\frac{\omega_k^2}{k^2}\right) + \Gamma_4 \,\frac{Z_L}{k^2} \frac{\vect{k}^2}{k^2}, \\
K_{CA} &= (\Gamma_S+\Gamma_4)\,\frac{Z_L}{k^2}\,\frac{\vect{q}\cdot\vect{k}}{k^2}\,\omega_k\,, \\
K_{CC} &= \Gamma_S\left(2\,\frac{Z_T}{k^2}  + \frac{Z_L}{k^2}\frac{\omega_k^2}{k^2} \right) \tilde{\omega}_q
	- \Gamma_4\,\frac{Z_L}{k^2}\frac{\vect{k}^2}{k^2}\,\tilde{\omega}_q\,.
\end{split}
\end{equation}

\section{Unquenching the gluon} \label{appUnqGluon}

In the Yang-Mills sector we solve the gluon DSE including the quark-loop
contributions shown in Fig.~\ref{fig:apprGluonDSE}, i.e.
\begin{align}
D_{\mu\nu}^{-1}(k) &= \left[D_{\mu\nu}^{que.}(k)\right]^{-1} - \sum_{f}^{N_f}\Pi_{\mu\nu}^f(k)\,, \label{eq:gluonDSEfullLoop} \\
 \Pi_{\mu\nu}^f(k) &= \frac{g^2 Z_2^f}{2}\is_p \Tr\left[ \gamma_\mu \,S^{f}(p) \,\Gamma^f_\nu(p,q;k) \,S^{f}(q)\right],   \nonumber
\end{align}
with the explicit flavor dependence indicated by the superscript $f$.
$D_{\mu\nu}^{que.}(k)$ denotes the quenched gluon propagator which is taken from
lattice calculations~\cite{Maas:2011ez}; the fitting procedure has been
discussed in Ref.~\cite{Fischer:2010fx}. $\Gamma^f_\nu(p,q;k)$ is given in Eq.~\eqref{eq:vertex1} but $\Gamma$ is evaluated for $p^2+q^2$ instead
of $k^2$ to ensure multiplicative renormalizability (cf.~Eq.~\eqref{PiT} below).
The lattice fits for the gluon dressing functions $Z_{{T},{L}}$ (cf.~Eq.~\eqref{eq:qProp}) are given by
\begin{equation}\label{eq:gluonFunction}
\begin{split}
Z_{{T},{L}}(k) &= \frac{x}{(x+1)^2} \bigg[
	\left(\frac{\hat{c}}{x + a_{{T},{L}}(T)}\right)^{b_{{T},{L}}(T)} \\
	& \qquad\qquad\quad\quad + x \left(\frac{\beta_0\,\alpha}{4\pi}\ln(1+x)\right)^\gamma\bigg],
\end{split}
\end{equation}
where $x=k^2/\Lambda^2$ and $a_{{T},{L}}(T)$, $b_{{T},{L}}(T)$ are temperature-dependent fit parameters.
In the ultraviolet, the logarithmic term leads to the perturbative running
with anomalous dimension $\gamma = \frac{-13N_c+4N_f}{22N_c-4N_f}$.
The temperature-independent parameters are $\hat{c}=5.87$ and $\Lambda=1.4$~GeV.
With a transition temperature of $T_c=277$~MeV for the quenched SU(3) theory, the
temperature-dependent parameters are given by
\begin{eqnarray}
a_{L}(t) &=& \twopartdef{0.595 - 0.9025\cdot t  + 0.4005\cdot t^2	  }{t<1}{3.6199\cdot t  - 3.4835}{t>1}, \nonumber\\
a_{T}(t) &=& \twopartdef{0.595 + 1.1010\cdot{}t^2			  }{t<1}{0.8505\cdot{}t - 0.2965}{t>1}, \nonumber\\
b_{L}(t) &=& \twopartdef{1.355 - 0.5741\cdot{}t + 0.3287\cdot{}t^2  }{t<1}{0.1131\cdot{}t + 0.9319}{t>1}, \nonumber\\
b_{T}(t) &=& \twopartdef{1.355 + 0.5548\cdot{}t^2			  }{t<1}{0.4296\cdot{}t + 0.7103}{t>1}
\end{eqnarray}
with $t := T/T_c$. Note that since this expression represents the quenched gluon
propagator it is independent of chemical potential, which enters the gluon DSE only through the quark loop, and $N_f=0$ in the anomalous dimension.

By contracting Eq.~\eqref{eq:gluonDSEfullLoop} with the projectors in Eq.~\eqref{eq:projTL} one arrives at the
equations for the transverse and longitudinal parts of the quark loop. Since we are using hard
momentum cutoffs in the numerical integration, these need to be carefully regularized to remove
quadratic divergencies without spoiling the Debye screening masses. This procedure is described
in the appendix of Ref.~\cite{Fischer:2012vc}. The resulting equations read
\begin{equation}\label{PiT}
    \Pi_{{T},{L}}(\vect{k}^2,0) = 2 g^2 Z_2  \is_p \frac{\Gamma(p^2+q^2)}{D(p)\,D(q)}\,K_{{T},{L}}
\end{equation}
for the contribution with lowest Matsubara frequency,
where $p$ and $q=p+k$ are the quark momenta in the loop and
\begin{equation*}
\begin{split}
K_{T} &= A(p)\,A(q)\,\Gamma_S \left( 3\,\frac{(\vect{p}\cdot\vect{k})^2}{\vect{k}^2} + 2\,\vect{p}\cdot\vect{k} - \vect{p}^2 \right), \\   K_{L} &= A(p)\,A(q)\left[\Gamma_S \left( 2\,\frac{\vect{p}\cdot\vect{k}\,\vect{k}\cdot\vect{q}}{\vect{k}^2} - \vect{p}\cdot\vect{q} \right)+ \Gamma_4 \,\vect{p}\cdot\vect{q} \right]  \\
& \quad + B(p)\,B(q)\,(\Gamma_4 - \Gamma_S) -C(p)\,C(q)\,(\Gamma_S+\Gamma_4)\,\tilde\omega_p^2\,.
\end{split}
\end{equation*}
The higher modes are accessed via
$\Pi_{{T},{L}}(\vect{k}^2,\omega_k)\rightarrow\Pi_{{T},{L}}(\vect{k}^2+\omega_k^2,0)$.
We use the same approximation in the quenched gluon propagator where we only have the
zero mode from the lattice. This procedure is surprisingly accurate, as has been verified
e.g. in \cite{Fischer:2010fx,Maas:2005hs}.

\section{Chiral condensate and definition of $T_C$ in a crossover region} \label{appCond}
The quark chiral condensate for a flavor $f$ is given by
\begin{equation}\label{eq:condensate}
\langle\bar{\psi}\psi\rangle_f = Z_2 Z_m  N_c  T\sum_n\int\frac{d^3p}{(2\pi)^3}\mathrm{Tr}_D\left[S_f(p)\right] ,
\end{equation}
where $Z_2$ is the quark wave function renormalization constant, $Z_m$ the quark mass renormalization constant, $N_c$ the number of colors and $\mathrm{Tr}_D$ indicates the Dirac trace.
Due to the nonzero quark mass the condensate is quadratically divergent and has to be regularized.
We consider two ways to define a (pseudo-) critical temperature that are both connected to the chiral condensate. %, which
The first one is the inflection point of the chiral condensate with respect to temperature,
\begin{equation}\label{eq:TCinflection}
T^{infl.}_C =  \smash{\displaystyle\max_{\forall T}} \left| \frac {\partial \langle\bar{\psi}\psi\rangle_f}{\partial T}\right|\, ,
\end{equation}
and the second one returns the maximum of the chiral susceptibility:
\begin{equation}\label{eq:TCchiralsusz}
T^\chi_C =  \smash{\displaystyle\max_{\forall T}} \left|\frac {\partial \langle\bar{\psi}\psi\rangle_f}{\partial m_f}\right|\,.
\end{equation}
Both definitions are independent of the regularization of the divergent part (finite bare masses).
While they give different results for $T_C$ in a crossover region,
they return the same position for
the critical end point.
\end{appendix}

\end{document}